# Paired Ru–O–Mo ensemble for efficient and stable alkaline hydrogen evolution reaction

HuangJingWei Li,[a,b,†] Kang Liu,[a,†] Junwei Fu,[a,*] Kejun Chen,[a] Kexin Yang,[a] Yiyang Lin,[a] Baopeng Yang,[a] Qiyou Wang,[a] Hao Pan,[c] Zhoujun Cai,[a] Hongmei Li,[a] Maoqi Cao,[a,e] Junhua Hu,[d] Ying-Rui Lu,[f] Ting-Shan Chan,[f] Emiliano Cortés,[g] Andrea Fratalocchi,[b] and Min Liu[a,*]

[a]State Key Laboratory of Powder Metallurgy, Hunan Provincial Key Laboratory of Chemical Power Sources, School of Physics and Electronics, Shenzhen Institute of Central South University, Central South University, Changsha 410083, P. R. China.
Email: fujunwei@csu.edu.cn, minliu@csu.edu.cn
[b]PRIMALIGHT, Faculty of Electrical and Computer Engineering, King Abdullah University of Science and Technology (KAUST), Thuwal, 23955-6900, Saudi Arabia.
[c]Department of Periodontics & Oral Mucosal Section, Xiangya Stomatological Hospital, Central South University, 72 Xiangya Road, Changsha, Hunan, PR China.
[d]School of Materials Science and Engineering, Zhengzhou University, Zhengzhou, 450002, P. R. China.
[e]School of Chemistry and Chemical Engineering, Qiannan Normal University for Nationalities, Duyun 558000, P. R. China.
[f]National Synchrotron Radiation Research Center, Hsinchu 300, Taiwan.
[g]Chair in Hybrid Nanosystems, Nano-Institute Munich, Faculty of Physics, Ludwig-Maxilimians-Universität München, München 80539, Germany.
† These authors contributed equally.

**Abstract**

Electrocatalytic hydrogen evolution reaction (HER) in alkaline media is a promising electrochemical energy conversion strategy. Ruthenium (Ru) is an efficient catalyst with a desirable cost for HER, however, the sluggish $H_2O$ dissociation process, due to the low $H_2O$ adsorption on its surface, currently hampers the performances of this catalyst in alkaline HER. Herein, we demonstrate that the $H_2O$ adsorption improves significantly by the construction of Ru–O–Mo sites. We prepared Ru/$MoO_2$ catalysts with Ru–O–Mo sites through a facile thermal treatment process and assessed the creation of Ru–O–Mo interfaces by transmission electron microscope (TEM) and extended X-ray absorption fine structure (EXAFS). By using Fourier-transform infrared spectroscopy (FTIR) and $H_2O$ adsorption tests, we proved Ru–O–Mo sites have tenfold stronger $H_2O$ adsorption ability than that of Ru catalyst. The catalysts with Ru–O–Mo sites exhibited a state-of-the-art overpotential of 16 mV at 10 mA $cm^{-2}$ in 1 M KOH electrolyte, demonstrating a threefold reduction than the previous bests of Ru (59 mV) and commercial Pt (31 mV) catalysts. We proved the stability of these performances over 40 hours without decline. These results could open a new path for designing efficient and stable catalysts.

# 1. Introduction

The production of hydrogen from $H_2O$ is considered one of the most promising approaches to meet the present and future demand for renewable electricity storage.[1-4] Currently, industrial hydrogen is mostly produced by electrolyzing $H_2O$ in alkaline solutions, involving hydrogen evolution reaction (HER) on cathodes and oxygen evolution reaction (OER) on anodes. One of the main issues that hamper the development of efficient systems lies in the slow HER kinetics in alkaline solutions, requiring the engineering of catalysts with sufficiently low overpotentials to drive the hydrogen reaction.[5-8]

In alkaline solutions, a fast HER kinetics requires the rapid dissociation of $H_2O$ and the subsequent hydrogen desorption.[9-13] Recently, ruthenium (Ru) has attracted attention as an ideal HER candidate with hydrogen (H*) adsorption energy (65 Kcal $mol^{-1}$) similar to the state-of-the-art Pt (62 Kcal $mol^{-1}$),[14-16] at a cost that is 25 times smaller than the cost of Pt.[17-19] However, the weak $H_2O$ adsorption and dissociation ability of Ru is a significant obstacle for fully exploiting the high intrinsic activity of this material in alkaline HER. The main issue is the weak interaction between the 4d orbitals of Ru and the 2p orbitals of the oxygen in $H_2O$ molecules.[20,21] The construction of properly paired M1–O–M2 (M means Metal) sites could be a promising strategy to overcome this problem. It has been reported that Pt–O–Ce and Pt–O–Pt sites have shown strong interactions between $O_2$ molecules and Pt sites to enhanced nine-fold catalytic efficiencies for CO oxidation.[22,23] Similarly, Cr–O–Ni reported good $H_2O$ dissociation kinetics and high neutral HER activity due to the enhanced interaction between $H_2O$ molecules and Cr–O–Ni sites.[24] Unfortunately, these systems are easily affected by oxidation of substrates, and no research succeeded in implementing M1–O–M2 sites for alkaline HER.

In this work, we design and demonstrate an alkaline HER $Ru/MoO_2$ composite with Ru–O–Mo sites, which has chemical stability and strong interaction with $H_2O$. Theoretical calculations performed with first-principle simulations predict that the engineering of Ru–O–Mo sites enhances $H_2O$ absorption.[25-29] Such enhancement is driven by the charge transfer from Ru sites to the O sites, brought by higher d-orbital of the surrounding Mo sites, which reduce the $H_2O$ dissociation barrier. Motivated by these theoretical results, we prepare $Ru/MoO_2$ with Ru–O–Mo sites by a hydrothermal process and pyrolysis treatment. The obtained catalyst shows an overpotentials as low as 16 mV at the current density of 10 mA cm$^{-2}$, with a Tafel slope of 32 mV dec$^{-1}$ in 1 M KOH solution and long-term HER stability over 40 hours.

## 2. Results and discussion

**Figure 1 and S1-2** shows the theoretical results on $Ru/MoO_2$ with Ru–O–Mo sites by density-functional theory (DFT) calculations. Calculated $H_2O$ adsorption energies show that the Mo sites of Ru–O–Mo have stronger $H_2O$ adsorption ability (2.06 eV) than Ru (0.35 eV) and $MoO_2$ (1.30 eV) (**Figure 1a and S3**). Next, we consider the effect of hydrogen binding energy (HBE) on water adsorption/dissociation. As shown in **Figure S4**, the HBE on Ru sites (–0.41 eV) of Ru-O-Mo is stronger than that on Mo sites (0.25 eV) of Ru–O–Mo, indicating that the Ru sites can adsorb hydrogen to further promote the $H_2O$ dissociation process. **Figure 1b** exhibits the charge density difference at the Ru–O–Mo sites. The 1.81 electrons are transferred from Ru to $MoO_2$ through the Ru–O–Mo sites, weakening the Mo–O bonds in $MoO_2$, enhancing the interaction of O in $H_2O$ molecules with the Mo sites.[26, 30] This process shorten the distance between the adsorbed $H_2O$ molecule and Ru–O–Mo sites of 2.19 Å, whose value is lower than pure $MoO_2$ (2.22 Å). These calculations demonstrate that Ru–O–Mo sites enhance the adsorption of $H_2O$ molecules.

**Figure 1c** reports the energy barriers of $H_2O$ dissociation and $H_2$ desorption on Ru–O–Mo sites. The energy barrier of $H_2O$ dissociation on Ru–O–Mo of 0.58 eV is significantly lower than that of Ru catalyst (0.81 eV), indicating that Ru–O–Mo sites achieve a better $H_2O$ adsorption and dissociation kinetics than pure Ru theoretically. The reduction of 0.23 eV in the $H_2O$ activation barrier, implying a massive reduction in the energy consumption. For $H_2$ desorption process, the Gibbs free energy of adsorbed *H ($\Delta G_{H^*}$) on the Ru–O–Mo sites (–0.27 eV) is much lower than that of Ru (–0.42 eV), demonstrating the easier desorption process on Ru–O–Mo sites.

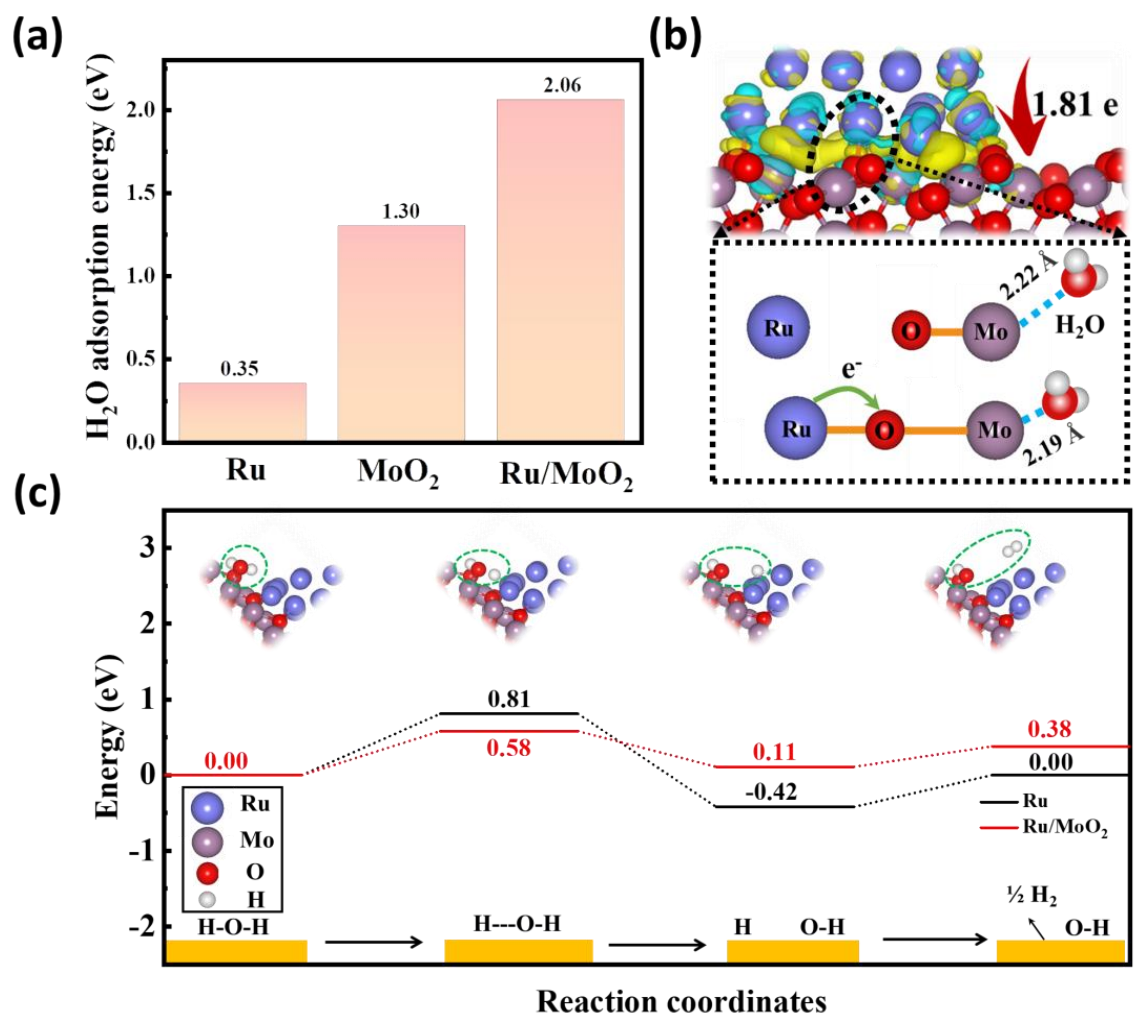

**Figure 1. The design principle and DFT calculations of catalysts.** (a) The calculated $H_2O$ adsorption energy of Ru, $MoO_2$, and Ru/$MoO_2$. (b) The charge density difference and Bader charge transfer at the interface of Ru/$MoO_2$, and the schematic diagram of charge transfer affecting $H_2O$ adsorption on interface Ru–O–Mo sites. The yellow and blue regions represent electron accumulation and depletion, respectively. (c) The calculated free-energy diagrams of $H_2O$ reduction to $H_2$ on the surface Ru and Ru/$MoO_2$ catalysts.

We experimentally prepare Ru/$MoO_2$ catalyst with paired Ru–O–Mo sites through a facile hydrothermal reaction with a pyrolysis process on a graphite carbon substrate (**Figure 2a**). Details on the manufacturing process are in the Supporting Information. **Figure S5** reports XRD pattern of obtained Ru/MoO$_2$ catalyst, showing the presence of MoO$_2$ (JCPDS 76-1807), and no obvious Ru XRD peaks due to the low Ru content.[18,31] Raman spectra confirm Mo–O bonds with sharp peaks at 479, 816, and 1000 cm$^{-1}$ [32,33], and a Ru–O bond peak at 646 cm$^{-1}$ (**Figure S6**).[34] According to the TG-DTA curves (**Figure S7**), each component's resulting content percentages in the catalyst are 21.26 wt% of Ru, 44.0 wt% of MoO$_2$, and 34.74 wt% of carbon.

We study the morphology of the obtained catalyst by transmission electron microscopy (TEM) (**Figure 2b** and **S8-10**). The small black spots are the Ru/MoO$_2$ catalyst, which is composed of Ru and MoO$_2$ nanoparticles with sizes ~2 nm and ~6 nm, respectively. A high-angle annular dark-field (HAADF) STEM image shows a clear interface of Ru–O–Mo sites. **Figure 2c** reports the corresponding lattice fringes, with interplanar spaces of 0.214 nm and 0.244 nm, representing the (002) plane of Ru and the (200) plane of MoO$_2$, respectively. **Figure 2d** presents energy-dispersive X-ray spectroscopy (EDS) mappings, which show uniform distribution of Ru, Mo, and O elements.

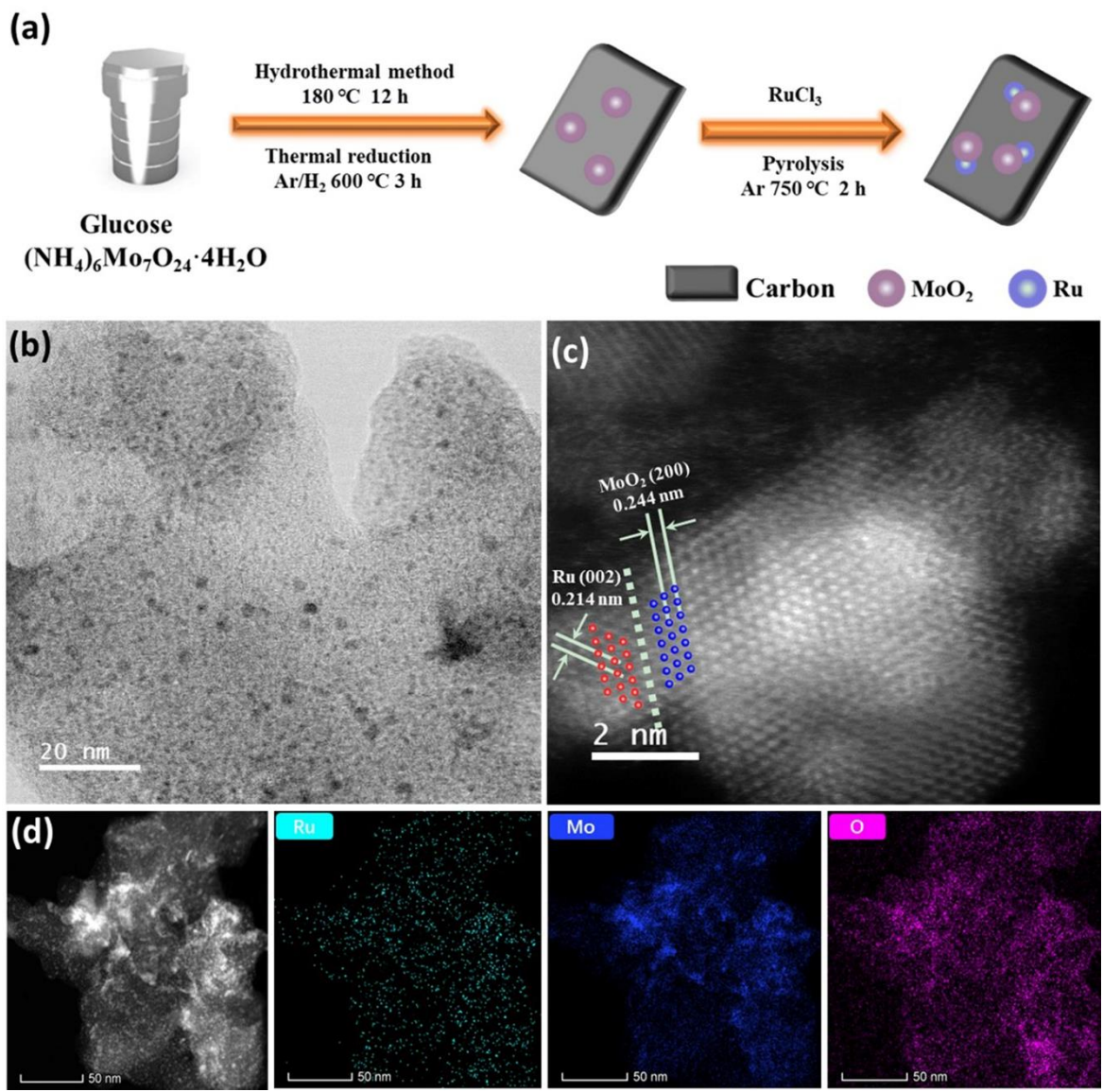

**Figure 2. Preparation process and structure characterizations of the catalyst.** (a) The schematic diagram of the preparation process of the Ru/MoO$_2$ catalyst. Typical TEM image (b), HAADF-STEM (c), and EDS mapping images (d) of the Ru/MoO$_2$ catalyst.

We further investigate the material structure of the catalyst by synchrotron X-ray absorption spectra.[35-37] The extended X-ray absorption fine structure (EXAFS) spectra and their wavelet transform analyses (**Figure 3a-b, S11, and Table S1**) showed that the obtained Ru/MoO$_2$ catalyst has clear Ru−O (1.5 Å) and Ru−Ru (2.4 Å) bonds.[38] No peak located at 3.2 Å of Ru−O−Ru bonding is observed in the catalyst, proving Ru–O–Mo sites form between Ru and MoO$_2$ nanoparticles. The positive shifts of Ru peak observed in both X-ray absorption near edge structure (XANES, **Figure 3c**) and X-ray photoelectron spectroscopy (XPS, **Figure 3d**) spectra of Ru/MoO$_2$ with Ru–O–Mo sites, further confirms the strong charge transfer from Ru to O. These results are consistent with the Bader charge analyses obtained by DFT calculations.

To prove the enhanced H$_2$O adsorption/dissociation ability on Ru/MoO$_2$, we carried out Fourier transfer infrared spectrum (FTIR) and H$_2$O adsorption tests. **Figure 3e** reports FTIR spectra, in which Ru/MoO$_2$ shows the strongest hydroxyl response signal among those of MoO$_2$ and Ru, in the range between 2900 cm$^{-1}$ ~ 3500 cm$^{-1}$.[39-43] In the H$_2$O adsorption tests (**Figure 3f and S12**), the higher current density difference between the cases with and without H$_2$O proves the larger H$_2$O adsorption/activation ability of Ru/MoO$_2$ with Ru–O–Mo sites,[44] which after normalization is between ~2 and ~10 times higher than that of MoO$_2$ and Ru, respectively. These experimental results confirm the DFT theoretical predictions on H$_2$O adsorption/dissociation energy.

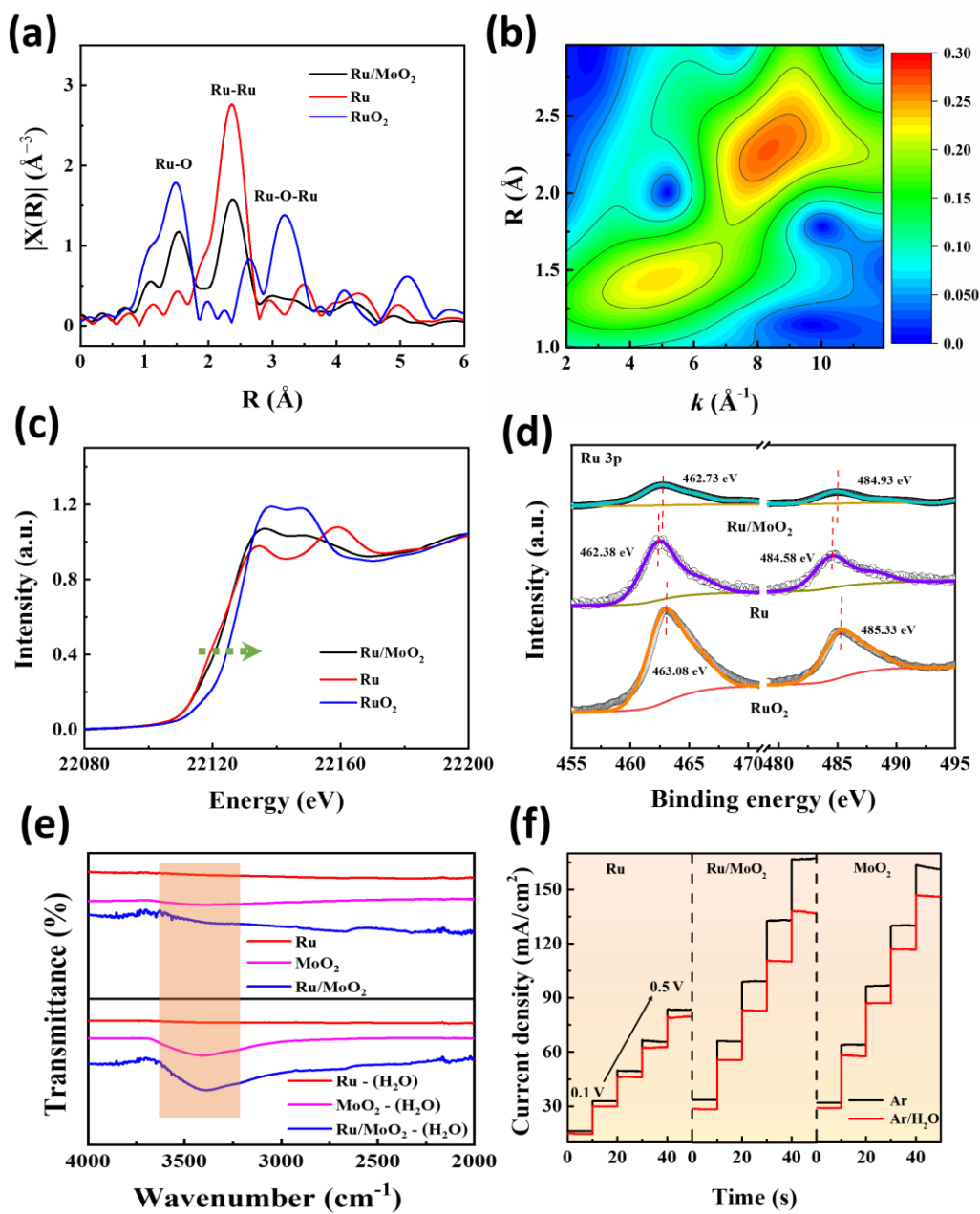

**Figure 3. Interface structure, charge transfer, and H₂O adsorption/dissociation ability.** (a) Extended X-ray absorption fine structure (EXAFS) spectra of Ru K-edge. (b) The Ru K-edge whole contour plots of wavelet transform (WT) of Ru/MoO₂ catalyst. (c) The enlarged X-ray absorption near edge structure (XANES) spectra of Ru K-edge. (d) High-resolution XPS spectra of Ru 3p of Ru, RuO₂,

and Ru/MoO$_2$. (e) The FTIR spectra of Ru, MoO$_2$, and Ru/MoO$_2$. (f) The H$_2$O adsorption sensor tests of Ru, MoO$_2$, and Ru/MoO$_2$.

We then assess the catalysts' electrocatalytic HER activities in the alkaline solution (1 M KOH) with a standard three-electrode system. As shown in **Figure 4a**, the linear sweep voltammetry (LSV) curves show that the Ru/MoO$_2$ with Ru–O–Mo sites has largely improved overpotential of 16 mV at 10 mA cm$^{-1}$, which is more than one order of magnitude lower than the one of MoO$_2$ (254 mV), and few times smaller than the ones of Ru (59 mV), and Pt (31 mV). The corresponding Tafel slope is 32 mV dec$^{-1}$, which shows a similar improvement over the 174, 64, and 39 mV dec$^{-1}$ possessed by Ru, MoO$_2$, and Pt, respectively (**Figure 4b**). The lower Tafel slope indicates faster HER kinetics of Ru/MoO$_2$ catalyst. The alkaline HER undergoes two important steps, one is H$_2$O adsorption/dissociation on the catalyst's surface (Volmer step: H$_2$O + * + e$^-$ → H* + OH$^-$), the other is hydrogen (H*) desorption (Tafel/Heyrovsky step: 2H* → H$_2$/H$_2$O + H* + e$^-$ → H$_2$ + OH$^-$).[12,45] Tafel slope is an important parameter to reflect the rate-determining step of HER. The 174 mV dec$^{-1}$ indicates the H$_2$O adsorption/dissociation (Volmer step) is the rate-determining step of Ru. The Tafel slope of 32 mV dec$^{-1}$ indicates an excellent H$_2$O adsorption/dissociation ability of Ru/MoO$_2$, which the rate-determining step transfers from H$_2$O adsorption/dissociation (Volmer step) to hydrogen (H*) desorption (Tafel/Heyrovsky step). Electrochemical impedance spectroscopy (EIS) is also use to investigate the electrode kinetics in HER.

The Nyquist plots of Ru/MoO$_2$ and other samples are shown in **Figure S13** and corresponding fitting element parament are shown in **Table S2**. The R$_{ct}$ of Ru/MoO$_2$ (24.59 Ω) is lower than Ru (35.53 Ω) and MoO$_2$ (39.81 Ω). Lower charge transfer resistance corresponds to a faster reaction rate of HER, indicating that Ru/MoO$_2$ is more active than other samples.

In addition to these benefits, Ru/MoO$_2$ with Ru–O–Mo sites show good durability in alkaline solutions. In the measurements performed over 5000 cycles, we observed only a small potential drop of 3 mV decayed at 10 mA cm$^{-2}$ (**Figure 4c**). Voltage-time curves (**Figure 4c, inset**) show that Ru/MoO$_2$ catalyst's potential is stable at 100 mA cm$^{-2}$ over 40 h. The SEM and XRD characterizations further verify that the morphology and structure of Ru/MoO$_2$ are well preserved after durability test. (**Figure S14 and S15**) **Figure 4d** and **Table S3**, reporting Tafel slopes and overpotentials at 10 mA cm$^{-2}$ with different Ru-based and Pt-based catalysts, demonstrate that the alkaline HER performance of the obtained Ru–O–Mo catalyst is superior to other reported catalysts.

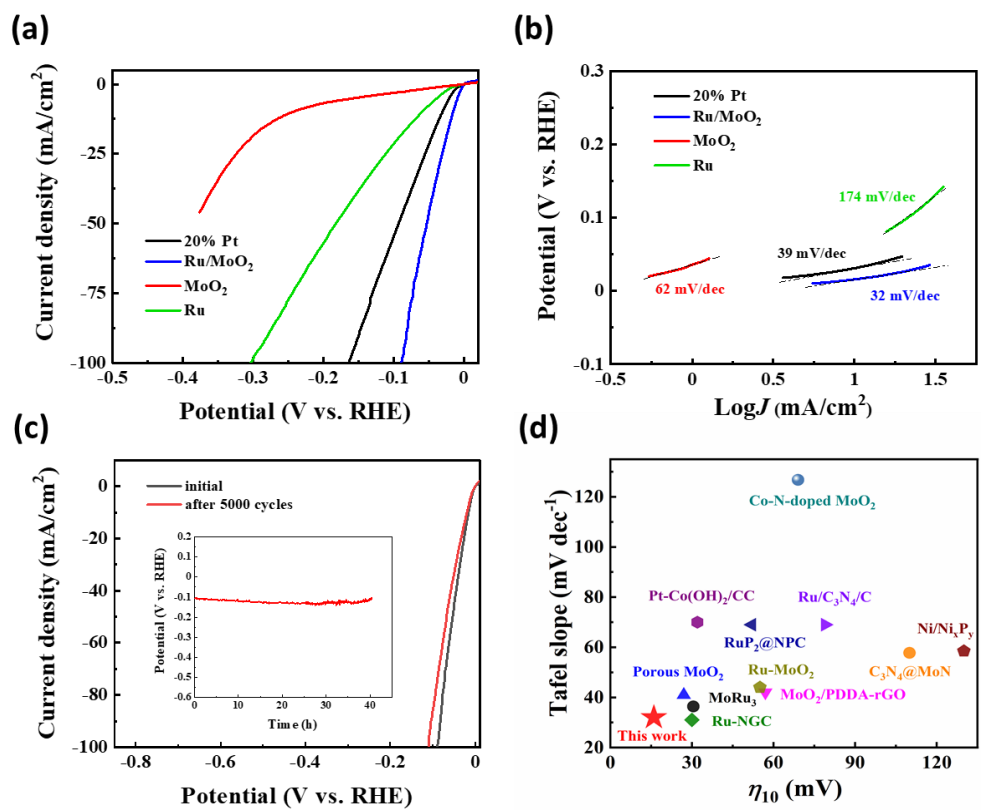

**Figure 4. The electrocatalytic HER tests in 1 M KOH solution.** (a) Linear scanning voltage (LSV) curves of commercial 20% Pt, Ru, MoO$_2$, and Ru/MoO$_2$ in 1 M KOH solution. (b) The corresponding Tafel plots are recorded in (a). (c) Linear scanning voltage curves of Ru/MoO$_2$ before and after 5000 cycles in 1 M KOH solution. The inset represents the voltage-time curve of Ru/MoO$_2$ at 100 mA cm$^{-2}$ for 40 h. (d) The comparisons of Tafel slopes and overpotentials at 10 mA cm$^{-2}$ with different Ru-based and Pt-based catalysts.

## 3. Conclusions

We designed and implemented an efficient catalyst for HER composed of Ru/MoO$_2$ with Ru–O–Mo sites. First-principle calculations showed that Ru–O–Mo sites have higher adsorption/dissociation for H$_2$O than conventional Ru catalysts. We then prepared Ru/MoO$_2$ with Ru-O-Mo sites, verifying the existence of Ru–O–Mo with HAADF-STEM images and EXAFS spectra. XANES and XPS experimentally proved the substantial interface charge transfer between Ru and MoO$_2$, while FTIR and H$_2$O adsorption tests demonstrated the enhanced H$_2$O adsorption ability of Ru–O–Mo sites. In a series of electrochemistry measurements, the catalyst implemented with this approach showed an overpotential of 16 mV (at 10 mA cm$^{-2}$), with stability over 40 hours in alkaline HER. This work can help in the design and implementation of highly-efficient alkaline HER catalysts for large scale, sustainable and economical hydrogen production from water splitting.

**Acknowledgments**

We thank the Natural Science Foundation of China (Grant No. 21872174 and U1932148), International Science and Technology Cooperation Program (Grant No. 2017YFE0127800 and 2018YFE0203402), Hunan Provincial Science and Technology Program (2017XK2026), Hunan Provincial Natural Science Foundation of China (2020JJ2041 and 2020JJ5691), Shenzhen Science and Technology Innovation Project (Grant No. JCYJ20180307151313532), The Hunan Provincial Science and Technology Plan Project (Grant No. 2017TP1001), The Fundamental Research Funds for the Central Universities of Central South University, and MOST 109-2113-M-213-002.